\documentclass[%
 reprint,
superscriptaddress,
nobibnotes,
 amsmath,amssymb,
 aps
prb,
]{revtex4-2}
\pdfoutput=1
\usepackage{graphicx,hyperref,amsthm,xcolor,soul,braket}% Include figure files
\usepackage[]{subcaption}
\usepackage{dcolumn}% Align table columns on decimal point
\usepackage{bm}% bold math
\usepackage{natbib}
\hypersetup{
    linkcolor=blue,
    citecolor=blue,
    filecolor=blue,
    urlcolor=blue,
    colorlinks=true
}

\DeclareMathOperator{\sgn}{sgn}
\newenvironment{nalign}{
    \begin{equation}
    \begin{aligned}
}{
    \end{aligned}
    \end{equation}
    \ignorespacesafterend
}
\begin{document}
%\preprint{APS/123-QED}
\title{Topological $d$-wave Superconductivity and Nodal Line-Arc Intersections in Weyl Semimetals}% Force line breaks with \\

\author{Gregory A. Hamilton }
\affiliation{ 
	Department of Physics, University of Illinois at Urbana-Champaign, Urbana, Illinois 61801, USA}
\affiliation{
Micro and Nanotechnology Laboratory, University of Illinois at Urbana-Champaign, Urbana, Illinois 61801, USA
}

\author{Moon Jip Park}
\affiliation{ 
Department of Physics, University of Illinois at Urbana-Champaign, Urbana, Illinois 61801, USA}
\affiliation{Department of Physics, Korea Advanced Institute of Science and Technology, Daejeon, 34141, Korea}

\author{Matthew J. Gilbert}
\affiliation{
Micro and Nanotechnology Laboratory, University of Illinois at Urbana-Champaign, Urbana, Illinois 61801, USA
}
\affiliation{
Department of Electrical and Computer Engineering, University of Illinois at Urbana-Champaign, Urbana, Illinois 61801, USA
}
\affiliation{Department of Electrical Engineering, Stanford University, Stanford, California 94305, USA}
\date{\today}

\begin{abstract}
Superconducting Weyl semimetals present a novel and promising system to harbor new forms of unconventional topological superconductivity. Within the context of time-reversal symmetric Weyl semimetals with $d$-wave superconductivity, we demonstrate that the number of Majorana cones equates to the number of intersections between the $ d $-wave nodal lines and the Fermi arcs. We illustrate the importance of nodal line-arc intersections by demonstrating the existence of locally stable surface Majorana cones that the winding number does not predict. The discrepancy between Majorana cones and the winding number necessitates an augmentation of the winding number formulation to account for each intersection. In addition, we show that imposing additional mirror symmetries globally protect the nodal line-arc intersections and the corresponding Majorana cones. 
\end{abstract}
\maketitle

Dirac and Weyl semimetals (WSMs) feature prominently in the study of topological materials alongside other semimetallic systems such as nodal line semimetals and semimetals with higher order degeneracies \cite{Sun2017,Bi2013,Fang2016,Bradlyn}. The low energy excitations of isolated gapless points (Weyl nodes) in the bulk Brillouin zone (BZ) of WSMs are Weyl fermions \cite{RevModPhys.90.015001,Yan2017,Weng2015}. Weyl nodes are two-fold  band degeneracies that are present in systems with either broken time-reversal symmetry (TRS) or inversion symmetry (IS)\cite{Weng2015,PhysRevLett.107.127205,Shuichi_Murakami_2007,PhysRevLett.109.066401}. In WSMs, open boundaries host topologically protected surface Fermi arcs connecting Weyl node projections of opposite chirality in momentum space \cite{RevModPhys.90.015001}. WSMs have been experimentally observed in a wide range of materials, most notably in the transition metal monopnictide class \cite{Weng2015,Huang2015a,Chiu2016b,Lv2015,Liu2015,Xu2015}. \par
The combination of WSMs and superconductivity is a powerful and robust platform for realizing novel topological phases of matter. In a topological superconductor (TSC), the quasiparticle spectrum has topologically protected, gapless Majorana modes that are essential to many topological quantum computing implementations \cite{Nayak2008}. Recent theoretical studies have primarily considered conventional ($s$-wave) or unconventional ($d$-wave) pairing in IS WSMs via bulk-doping \cite{Sato2017,Lu2015,Li2018,Bednik2015,Hosur2014a} or proximity effects \cite{Sato2017,Wang2017,Khanna1954,Kim2016a,Chen2016}. In the case of TRS WSMs, TRS superconducting pairing between Weyl nodes of opposite momentum and equal chirality opens a bulk superconducting gap so long as the pairing potential does not vanish at the Weyl nodes \cite{Meng2012}. In fully gapped superconductors, the sign of the pairing potential, combined with Fermi surfaces possessing non-zero chirality, define the topological invariants for classes of 3D TRS TSCs \cite{Qi2010}. Within the weak superconducting pairing limit, the relevant topological invariant is the winding number, given by 
 \begin{nalign}\label{eq:winding}
 	N_{w} = \frac{1}{2}\sum_{s}C_{s}\sgn{\Delta_{s}}.
 \end{nalign} 
Here $C_{s}$ denotes the first Chern number on the $s$-th disconnected Fermi surface, and $\Delta_{s}$ denotes the effective pairing gap on the $s$-th Fermi surface. Eq. \eqref{eq:winding} determines the number of protected gapless modes along an open boundary. Furthermore, Eq. \eqref{eq:winding} indicates sign changes in the pairing potential are an important ingredient to realize TSC, as a constant pairing potential implies, by the Nielsen-Ninomiya theorem, a trivial winding number \cite{NIELSEN1981219}. However, unconventional nodal superconductivity, such as TRS $d$-wave superconductivity, naturally possesses sign changes in the pairing potential. Therefore, TRS WSM with $d$-wave superconductivity is a natural candidate for TSC.

%
%However, Eq. \eqref{eq:winding} gives no insight into the total number and location of Majorana modes in the surface Brillouin zone. Furthermore, Eq. \eqref{eq:winding} cannot account for (non)symmorphic crystalline symmetries, often present in WSMs, that can substantially alter the topological classification \cite{Chiu2013,Schnyder,Sato2017,Shiozaki2016}. Addressing these limitations prompts a more careful study of the interplay between the WSM topology and the superconducting pairing potential to determine the presence of TSC. 
In this work, we study TSC in TRS WSMs with  $d$-wave superconductivity. We demonstrate that intersections between the Fermi arcs and the nodal lines in the pairing potential naturally host Majorana cones. Interestingly, we show that additional, locally stable Majorana cones occur at the nodal line-arc intersections (NAIs) that the winding number cannot account for. Confronting this winding number limitation prompts a more careful study of the interplay between the WSM topology and the superconducting pairing potential to determine the presence of TSC. We address the limitation by recasting Eq. \eqref{eq:winding} to give an alternative definition of the winding number as a function of NAIs. Motivated by the mirror symmetries present in TRS WSMs such as TaAs and TaP \cite{Huang2015a,Lv2015,Liu2015,Xu2015}, we consider the addition of mirror symmetries and determine the augmented topological classification that crucially depends upon the nature of NAIs. 

We begin with a phenomenological Bogoliubov de-Gennes (BdG) Bloch Hamiltonian that describes TRS WSMs with $d$-wave pairing
\begin{nalign}\label{eq:phenHam}
	{h}_{BdG}(\bm{k}) = \begin{pmatrix}
		h_{0}(\bm{k}) & \Delta(\bm{k}) \\
		\Delta^{\dagger}(\bm{k}) & -h_{0}^{\ast}(-\bm{k})
	\end{pmatrix}.
\end{nalign}
In this formulation, $h_{BdG}(\bm{k})$ acts on the Nambu spinor $\Phi_{\bm{k}} = (\Psi_{\bm{k}}, \Psi_{-\bm{k}}^{\dagger})^{T}$ \cite{Bogoljubov1958,Sato2017}, and $\Delta(\bm{k})$, $ h_{0}(\bm{k})$ denote the $ d $-wave pairing matrix and the TRS WSM Bloch Hamiltonian, respectively. The full Hamiltonian written in terms of Eq. {\eqref{eq:phenHam}} commutes with both the TRS and particle-hole symmetry (PHS) operators. In terms of $h_{BdG}(\bm{k})$, these symmetries are written as $U\mathcal{K}$, where $U$ is a unitary operator and $\mathcal{K}$ is complex conjugation. $U_{T} := is_{y}, U_{C} := i\tau_{x}$ act on $h_{BdG}(\bm{k})$ as 
\begin{nalign}
U_{T}h_{BdG}^{\ast}(-\bm{k})U_{T}^{-1} = h_{BdG}(\bm{k}),  \\
-U_{C}h_{BdG}(-\bm{k})U_{C}^{-1} = h_{BdG}(\bm{k}),
\end{nalign} 
where $\bm{\tau}, \bm{s}$ denote the Pauli spin vectors in the Nambu and spin spaces, respectively. Due to TRS, a Weyl node and its TRS partner have the same chirality, thereby ensuring a gapped bulk quasiparticle spectrum in the presence of $d$-wave superconductivity, so long as the Weyl nodes do not occur at nodal lines \cite{Meng2012}. TRS further prohibits a complex phase in the pairing potential \cite{Hosur2014a}, implying that sign changes accompany nodal lines in the superconducting order parameter. \par 
\begin{figure}
	\includegraphics[trim=5cm 0 5cm 0,clip,width=.8\linewidth]{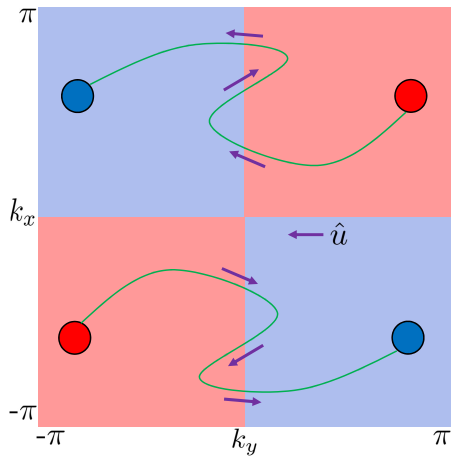}
	\caption{Schematic of Fermi arcs (green) in a TRS WSM intersecting nodal lines in the $d_{xy}$-pairing potential at an open boundary. The blue (red) points depict surface Weyl node projections of negative (positive) chirality. The same coloring scheme indicates the different signed regions of the $d_{xy}$-pairing potential. The tangential arrows $\hat{u}_{i}$ depict chiral flow from the positive to negative chirality Weyl node projections.}\label{fig:windingFormula}
\end{figure}
We now consider an open boundary in the $\hat{z}$-direction and assume that the superconducting pairing potential has no dependence in $k_z$. Fig. \ref{fig:windingFormula} depicts a hypothetical configuration of Fermi arcs connecting Weyl node projections at the open $\hat{z}$ boundary where the nodal lines of a $d_{xy}$-pairing potential are superimposed. By Eq. \eqref{eq:winding}, the winding number is $N_{w}= 2$ for this configuration of Weyl nodes. In general, a non-zero $N_{w}$  implies that, for any normal phase Fermi arc configuration, at least $|N_{w}|$ intersections between Fermi arcs and nodal lines in the pairing potential must occur, as shown in Fig. {\ref{fig:windingFormula}}.  

To relate the NAIs and the bulk winding number, we introduce an orientation of the Fermi arcs that starts at the positive chirality Weyl node projection and ends at the negative chirality Weyl node projection. The oriented NAIs allows us to recast the winding number as 
\begin{nalign}\label{eq:recastedWinding}
	N_{w}= -\sum_{i}\sgn(\nabla \Delta_{i}\cdot \hat{u}_{i}),
\end{nalign}
where $i$ sums over all the NAIs in the BZ, $\nabla \Delta_{i}$ is the gradient of the pairing gap function at the intersection, and $\hat{u}_{i}$ is a unit vector tangent to the Fermi arc at the intersection and pointing along the orientation of the arc.

To illustrate the importance of NAIs in understanding the nature of the surface physics, we note that, by definition, $\Delta(\bm{k}) = 0$ at NAIs. At these momenta, $h_{BdG}(\bm{k})$ decouples into particle and hole sectors 
\begin{nalign}
h_{BdG}(\bm{k}) \cong h_{0}(\bm{k})\oplus -h_{0}^{\ast}(-\bm{k}).
\end{nalign} 
A Fourier transform in the open boundary direction $\hat{z}$ leaves $h_{BdG}(\bm{k}_{\parallel},z)$ block-diagonal, where $\bm{k}_{\parallel} = (k_{x},k_{y})$. If $u(\bm{k}_{\parallel},z)$ is a zero-energy surface eigenstate of $h_{0}(\bm{k}_{\parallel},z)$, then $v(\bm{k}_{\parallel},z):=u^{\ast}(-\bm{k}_{\parallel},z)$ is a zero-energy surface eigenstate of $h_{0}^{\ast}(-\bm{k}_{\parallel},z)$. Similarly, by TRS, $Th_{0}^{\ast}(-\bm{k}_{\parallel},z)T^{-1}= h_{0}(\bm{k}_{\parallel},z)$; thus, $Tu(\bm{k}_{\parallel},z)T^{-1} = v(\bm{k}_{\parallel},z)$, which implies $u^{\ast}(-\bm{k}_{\parallel},z) = v(\bm{k}_{\parallel},z)$. Therefore, the Bogoliubov transformation, 
\begin{nalign}\gamma_{\bm{k}_{\parallel},z} = u(\bm{k}_{\parallel},z)\Psi_{\bm{k}_{\parallel},z} + v(\bm{k}_{\parallel},z)\Psi_{-\bm{k}_{\parallel},z}^{\dagger},\end{nalign} 
which is the eigenstate of the surface Hamiltonian, satisfies $\gamma_{\bm{k}_{\parallel},z}^{\dagger} = \gamma_{-\bm{k}_{\parallel},z}$. This is precisely the criterion for a Majorana operator under open boundary conditions \cite{Bjornson2015}. The linear dispersion of both the Fermi arc and the pairing potential at the nodal line ensures Majorana surface cones occur at every NAI. \par 
 \begin{figure}
 \includegraphics[width=\linewidth,trim=.21cm 3.2cm 0 3.55cm, clip]{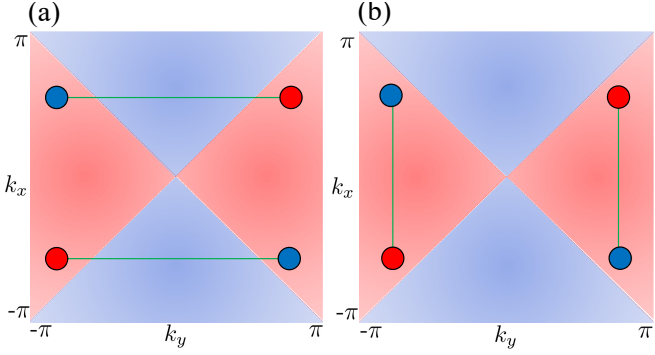}
 \caption{ (a-b) Allowed Fermi arc connectivities under a fixed configuration of Weyl nodes on an open boundary with $d_{{x^2}-{y^2}}$-pairing potential superimposed on the plot. Here the winding number is trivial, as the minimal number of NAIs is zero (b).}
 \label{fig:pairingGeometry}
 \end{figure}
Eq. {\eqref{eq:recastedWinding}} makes clear that the total number of NAIs and thus Majorana cones is not  equivalent to the number of topological Majorana cones. Thus, additional Majorana cones beyond those required by the winding number occur as pairs in the quasiparticle spectrum. Due to chirality, these additional unprotected Majorana gap pairwise when brought together in momentum space by an adiabatic deformation of the Fermi arc configuration. Therefore, if the number of NAIs is preserved, then the accidental Majorana cones are stable in the clean limit. 

The inclusion of mirror symmetries such as those present in TaAs and TaP fundamentally alters the topological structure of the BdG Hamiltonian, thus changing the number of topologically protected Majorana modes at NAIs. To illustrate how the topological protection at NAIs changes under mirror symmetry, we consider a tight-binding TRS WSM Hamiltonian \footnote{ This Hamiltonian is inspired by Ref. \cite{Wang2017}} 
\begin{equation}
	\label{eq:normalH}
	% \begin{aligned}
	% 	h_{0}(\bm{k}) & = (m_{z} + A\left(3-\cos k_{0}\cos 2k_{x} - \cos k_{y} - \cos k_{z}\right))\sigma_{z} 
	% \\
	%  &+ B\sin k_{0}\cos 2k_{x}\sigma_{x} + B \cos k_{0}\sin 2k_{x}\sigma_{x}s_{y}  
	% \\
	%  & + B \sin k_{z} \sigma_{y}+ A \sin k_{0} \sin 2k_{x}\sigma_{z}s_{y} \\
	%  & + \lambda_{x}s_{y} + \lambda_{y}\sigma_{y}s_{x},
		\begin{aligned}
		h_{0}(\bm{k}) & = (m_{z} + A\left(3-\cos k_{0}\cos 2k_{x} - \cos k_{y} - \cos k_{z}\right))\sigma_{z} \\
	 &+ B\sin k_{0}(\cos 2k_{x}((1-\lambda)+ \lambda \cos2k_{y}))\sigma_{x} \\
	 & + B \cos k_{0}\sin 2k_{x}\sigma_{x}s_{y}  + B \sin k_{z} \sigma_{y} \\
	 & + A \sin k_{0} \sin 2k_{x}\sigma_{z}s_{y}  + \frac{\lambda}{2}(\sin k_{y}s_{x}-\sin k_{x}s_{y}),
	\end{aligned}
\end{equation} 
where $\sigma_{i}$ are the Pauli matrices that act on orbital space. The parameters $A , \, B,$ and $\lambda$ break IS that, by definition, sends $\bm{k}\to -\bm{k}$. The last term is a Rashba coupling term that alters the connectivity of the Fermi arcs when $\lambda$ is changed, as described below. $h_{0}(\bm{k})$ also respects the mirror symmetries $M_{x/y} := is_{x/y}$. The superconducting pairing matrix takes the form
\begin{nalign}
	\Delta(\bm{k}_{\parallel}):= i\Delta_{\alpha}(\bm{k}_{\parallel})\sigma_{0}s_{y},
\end{nalign} 
where $\Delta_{\alpha}$ denotes the two types of $d$-wave pairing gap functions, either $\Delta_{xy}$ or $\Delta_{x^{2}-y^{2}}$, given by
\begin{nalign}\label{eq:formsPairing}
	\Delta_{xy}(\bm{k}_{\parallel}) = \Delta_{0}\sin k_{x} \sin k_{y}, \\
	\Delta_{x^{2}-y^{2}}(\bm{k}_{\parallel}) = \Delta_{0}(\cos k_{x} - \cos k_{y}).
\end{nalign} 
Since the pairing potential vanishes at nodal lines for both intra- and inter-orbital pairing, we include only intra-orbital pairing and set the magnitude to be $\Delta_{0}=1$.

In the BdG Hamiltonian, $h_{BdG}(\bm{k})$ satisfies $M_{i}$ symmetry if
\begin{nalign}
M_{i}h_{BdG}(k_i,\bm{k})M_{i}^{-1} = h_{BdG}(-k_{i},\bm{\tilde{k}}),
\end{nalign} 
where $\bm{\tilde{k}}$ denotes the momenta unaffected by $M_{i}$. For $d_{x^{2}-y^{2}}$-pairing, the mirror symmetries are given by $M_{x}= i\tau_{z}s_{x}, \, M_{y}= is_{y}$, while for $d_{xy}$-pairing the mirror symmetries are written as $M_{x}= is_{x}, \, M_{y}= i\tau_{z}s_{y}$. The difference in the form of $M_{i}$ between the two types of $d$-wave pairing arises from a $U(1)$-gauge choice in the hole sector of $h_{BdG}(\bm{k})$ \cite{Sato2017}. Specifically, if $M_{i}^{0}$ acts on $h_{0}(\bm{k})$, then $M_{i}^{BdG} = M_{i}^{0}\tau_{0}$ or $M_{i}^{BdG} = M_{i}^{0}\tau_{z}$, depending on whether or not the pairing potential changes sign under the mirror reflection.
 
To determine the topological classification under $M_{i}$ symmetry, we utilize the minimal Dirac Hamiltonian method, briefly summarized in Appendix \ref{app:MDH}, which involves analyzing the existence or non-existence of additional symmetry-preserving extra mass terms in topological Dirac Hamiltonians \cite{Chiu2013,Shiozaki2014,Schnyder}. The extra mass terms depend upon indices $\eta_{T}^{i},\eta_{C}^{i}$ that respectively satisfy \footnote{ In contrast to the convention in Ref. \cite{Chiu2013}, we do not force $M_{i}$ to be Hermitian.}
\begin{nalign}\label{eq:commRel}
M_{i}U_{T}\mathcal{K} = -\eta_{T}^{i}U_{T}\mathcal{K}M_{i}, \\ 
M_{i}U_{C}\mathcal{K} = -\eta_{C}^{i}U_{C}\mathcal{K}M_{i}.
\end{nalign}

\begin{figure*}[t!]
	\centering
	\includegraphics[width=1\textwidth,trim=0cm 3cm 0cm 3.6cm, clip]{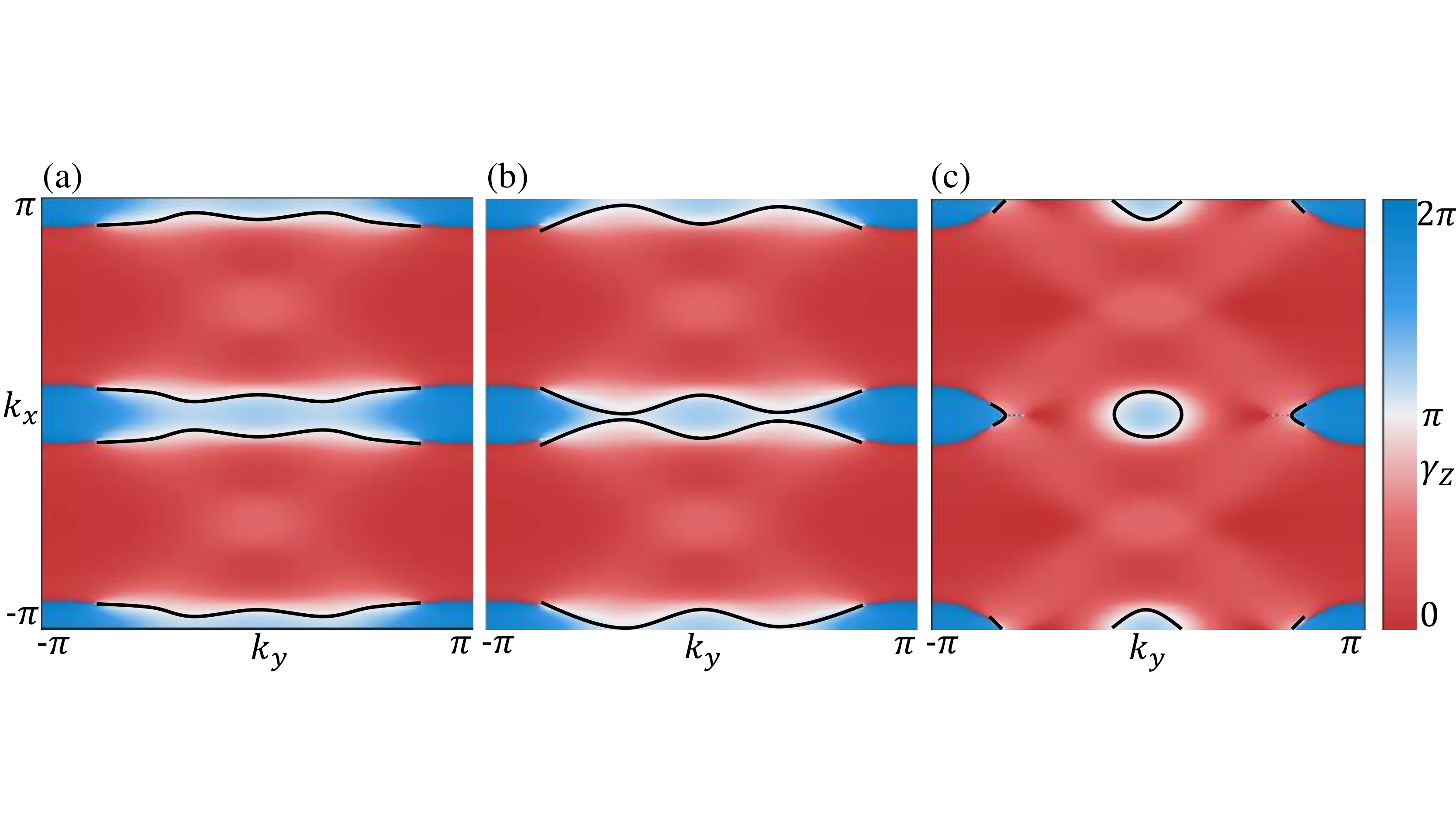}
	\caption{The changing of the Fermi arc connectivity (black lines) in the presence of $M_{y}$ symmetry. The initial Fermi arc configuration (a) ($\lambda =0$) is
		%in the normal phase. 
		shifted in a mirror-symmetric manner (b) ($ \lambda = 0.1 $), such that  
		%causing a redistribution of the arc connectivity. 
		the mirror Chern number forces a dangling Dirac cone to appear once the Weyl node connectivity is changed (c) $ (\lambda=0.5) $. In this figure, we have chosen $m_{z}= A/(-2+\sqrt{2}), \, k_{0} = \pi/4, \ A =B =2$. }\label{fig:mirrorDanglingDirac}
\end{figure*}

There is a profound difference in the topological classification of the BdG Hamiltonian depending on the values of  $\eta_{T}^{i},\eta_{C}^{i}$. To illustrate the difference, we begin with $d_{x^{2}-y^{2}}$-pairing that corresponds to $\eta_T=-1, \eta_C=-1$. We find that $d_{x^{2}-y^{2}}$-pairing is always topologically trivial under both $M_{x}$ and $M_{y}$ symmetries. We may understand the emergence of trivial topology in the presence of $d_{x^{2}-y^{2}}$-pairing by considering a bulk Weyl node situated away from any high symmetry points in the BZ and its TRS partner of the same chirality. Upon including either $M_{x}$ or  $M_{y}$ symmetry, both Weyl nodes now have a partner of opposite chirality, as depicted in Fig. \ref{fig:pairingGeometry}(a). Neighboring nodes of opposite chirality always occur in regions of the same sign of the pairing potential. Eq. \eqref{eq:recastedWinding} then implies that the contribution of each pair of neighboring nodes to the winding number vanishes. Furthermore, for $d_{x^{2}-y^{2}}$-pairing there always exists a Fermi arc configuration wherein neighboring Weyl nodes in the same sign of pairing potential may be connected, thus implying the winding number vanishes, as shown in Fig. \ref{fig:pairingGeometry}(b).\par 
In contrast to $d_{x^{2}-y^{2}}$-pairing, $d_{xy}$-pairing, corresponding to $\eta_{T} = -1, \ \eta_{C} = 1$, supports a $\mathbb{Z} \oplus \mathbb{Z}$ classification \cite{Chiu2013}. The first $\mathbb{Z}$ index corresponds to the winding number as given in Eq. \eqref{eq:recastedWinding}, while the latter index corresponds to the mirror strong index, $N_{M_{i}\mathbb{Z}} $. While the topological classification is extended under $d_{xy}$-pairing to include two invariants, the number of protected Majorana modes is given by $\max(|N_{M_{i}\mathbb{Z}}|,|N_{w}|)$ \cite{Chiu2013}. Assuming a generic $h_{0}(\bm{k})$ respects $M_{y}$ symmetry, we block diagonalize $h_{BdG}(\bm{k})$ into $h_{BdG,+i}(\bm{k})\oplus h_{BdG,-i}(\bm{k})$, where $\pm i$ are the eigenvalues of $M_{y}$. The mirror strong index is then given by \cite{Chiu2013} 
\begin{nalign}\label{eq:MSI}
N_{M_{y}\mathbb{Z}} = \sgn(M\mathbb{Z}^{BdG}_{0} - M\mathbb{Z}^{BdG}_{\pi})(|M\mathbb{Z}^{BdG}_{0}|-|M\mathbb{Z}^{BdG}_{\pi}|).
\end{nalign} 
 Here the mirror Chern number, $M\mathbb{Z}_{i}^{BdG}$, is defined as 
\begin{nalign}
	M\mathbb{Z}_{i}^{BdG} = \frac{1}{2}(C^{+,BdG}_{i} - C^{-,BdG}_{i}),
\end{nalign} 
where $C_{i}^{+,BdG}$ denotes the Chern number for $h_{BdG,+}$ on the surface $k_{y}=i$ in the bulk Brillouin zone \cite{Teo2008a}. \par

Turning to the tight-binding model given in Eq. \eqref{eq:normalH}, we choose the parameters $A, \, B,\, k_{0}$, and $ m_{z}$ such that eight zero-energy Weyl nodes are located in the BZ as shown in Fig. \ref{fig:mirrorDanglingDirac}(a). In Fig. \ref{fig:mirrorDanglingDirac}(a), we show the Zak phase corresponding to $h_{0}(\bm{k})$ ($\lambda=0$) along the $\hat{z}$ boundary, defined to be $\gamma^{Z}(\bm{k}_{\parallel}) = \oint \text{d}k_{z}A(\bm{k})$, where $A(\bm{k})$\ is the Berry connection. The regions where the Zak phase equals $\pi$ denote the locations of the four Fermi arcs, while the vorticity at the Weyl node projections indicates the sign of the chirality of the bulk Weyl node \cite{Zak1989,Kim2016b}. Including $d_{xy}$-pairing and using Eq. \eqref{eq:recastedWinding}, we find $N_w=0$, while, in contrast, $|N_{M_{y}\mathbb{Z}}|=4$. We describe the details of this calculation in Appendix \ref{app:mirrorChern}.
 The non-zero value of the mirror strong index is evident from the geometrical configuration of the Fermi arcs. Since the pairing potential vanishes along the mirror plane, the mirror strong index is simplified to $|N_{M_{y}\mathbb{Z}}| =|M\mathbb{Z}_{0}|$, where $M\mathbb{Z}_{0}$ denotes the mirror Chern number of $h_{0}(\bm{k})$. As {$M_{y}$} symmetry conserves the Chern number in each mirror sector, Fermi arcs with opposite chirality cannot hybridize to produce a gap at {$k_y=0$}. Therefore, the number of zero modes intersecting {$k_y=0$} must be preserved. Since there are four Fermi arcs crossing $k_y=0$ (two Fermi arcs per mirror sector, as shown in Fig. {\ref{fig:mirrorDanglingDirac}(a)}), the mirror strong index is $|N_{M\mathbb{Z}}|=4$. Correspondingly, the four locations where the Fermi arcs cross $k_{y}=0$ exhibit Majorana modes.\par 

We further elucidate the relation between NAIs and $ N_{M_{y}\mathbb{Z}} $ by considering mirror-symmetric deformations of the Fermi arcs that change the connectivity of the Weyl nodes,
 as shown in Fig. \ref{fig:mirrorDanglingDirac}. Noting that the mirror planes and nodal lines coincide, the particle and hole sectors decouple and we may restrict our analysis to the normal phase Hamiltonian. Starting from Fig. \ref{fig:mirrorDanglingDirac}(a), we increase $\lambda$ while preserving $M_{y}$ symmetry, thereby changing the Fermi arc configuration to that shown in Fig. \ref{fig:mirrorDanglingDirac}(b)  ($ \lambda =0.1 $) and, finally, arriving at Fig. \ref{fig:mirrorDanglingDirac}(c) $ (\lambda  = 0.5) $. Fig \ref{fig:mirrorDanglingDirac}(c) shows the final altered Fermi arc configuration after reconnecting the Weyl nodes where, despite changing the Weyl node connectivity, the number of zero modes crossing $k_{y}=0$ is conserved by the creation of a disconnected Dirac cone \footnote{The dangling Dirac cones are distinct from the dangling Dirac cones related to the $\mathbb{Z}_{2}$ weak indices derived from TRS in \cite{Lau2017}, as the number of Dirac cones present in this case are protected by a $\mathbb{Z}$ rather than a $\mathbb{Z}_{2}$ invariant.}. These dangling Dirac cones ensure that the $\mathbb{Z}$ mirror Chern number remains unchanged in the normal phase. The coincidence of nodal lines and mirror planes in $d_{xy}$-pairing ensures the topological protection from the mirror symmetry carries through from the normal phase to the superconducting phase. Moreover, in the superconducting phase, after the deformation four surface Majorana cones appear at the NAIs along $ k_{x} = 0 $. Thus, the mirror symmetry ensures that the number of NAIs are preserved even when the Weyl node connectivity is changed.  \par 
In conclusion, we considered a TRS WSM with {$d$}-wave superconductivity and analyzed the resulting topological classification and gapless surface modes. We demonstrated, both analytically and numerically, the existence of locally stable Majorana cones at NAIs that the DIII winding number does not predict. Consequently, we provided an augmentation of the winding number
formalism that specifies both the number and location of all Majorana cones that occur along an open boundary in the $\hat{z}$-direction. Given the mirror symmetries inherent to many experimentally observed TRS WSM, such as TaAs, TaP, and NbAs, we further analyzed how the topological classification changes when we incorporate the mirror symmetries of both the TRS WSM and the {$d$}-wave pairing potential. We find that the mirror symmetries of the two {$d$}-wave pairing potentials considered, {$d_{xy}$}- and {$d_{x^{2}-y^{2}}$}-pairing, give rise to drastically different topological classifications. The mirror symmetry under {$d_{xy}$}-pairing protects surface Majorana cones, even when the Fermi arc connectivity is changed in a mirror symmetric manner, while the mirror symmetry of {$d_{x^{2}-y^{2}}$}-pairing renders the system topologically trivial. Our results further extend predictions of topological superconductivity in TRS WSM, and highlight the crucial roles unconventional superconductivity, crystalline symmetries, and Fermi arcs play in understanding these exotic systems.

\acknowledgments
G.A.H, M.J.P, and M.J.G. acknowledge financial support from the National Science Foundation (NSF) under Grant No. DMR-1720633. M.J.P. is supported by the BK21 plus program, KAIST, and the National Research Foundation Grant NRF2017R1A2B4008097. M.J.G. acknowledges support from the NSF under CAREER Award ECCS-1351871 and the Office of Naval Research (ONR) under grant N00014-17-1-3012. G.A.H acknowledges fruitful conversations with Y. Kim, M. R. Hirsbrunner, T. M. Philip, and B. Basa.
\appendix
\section{Numerical Mirror Chern Number Calculation}\label{app:mirrorChern}
In this section, we review a numerical approach for finding the Chern number for a mirror sector. In doing so, we closely follow the notation given in \cite{Fukui2005}. 
 
We begin with a generic $n$-band Bloch Hamiltonian, where we label $\ket{n(k)}$ as the (normalized) $n$-th band wavefunction. We define a ``link'' variable \cite{Fukui2005} as 
\begin{nalign}
U_{\mu}(\bm{k}_{l}) = \braket{n(\bm{k}_{l})|n(\bm{k}_{l}+\hat{\mu})}/\mathcal{N}_{\mu}(\bm{k}_{l}),
\end{nalign} 
where $\hat{\mu}$ is a lattice unit vector, and $\bm{k}_{l}$ denotes a momentum in the (discretized) lattice. Here $\mathcal{N}_{\mu}(\bm{k}_{l}) = |\braket{n(\bm{k}_{l})|n(\bm{k}_{l}+\hat{\mu})}|$. Then the Chern number (explicitly summing over occupied bands) is given by 
\begin{nalign}
C_{1} = \frac{1}{2\pi i}\sum_{n\in occ}\sum_{l}F^{n}_{ij}(\bm{k}_{l}),
\end{nalign} 
where the $n$-th band field strength $F^{n}_{ij}(\bm{k}_{l})$ is given by 
\begin{nalign}
F_{ij}(\bm{k}_{l}) = \ln U_{i}(\bm{k}_{l})U_{j}(\bm{k}_{l}+ \hat{i})U_{i}(\bm{k}_{l}+\hat{j})^{-1}U_{j}(\bm{k}_{l})^{-1}.
\end{nalign}
 Here $\hat{i}, \hat{j}$ label unit lattice vectors, and $F_{ij}(\bm{k}_{l})$ is defined within the principal branch of the logarithm \cite{Fukui2005}.
 In the case of degeneracies in the occupied bands, the non-Abelian connection must be used, and the ``link'' variable is replaced with 
\begin{nalign}
U_{\mu}(\bm{k}_{l}) = \det( \braket{\psi(\bm{k}_{l})|\psi(\bm{k}_{l}+\hat{\mu})} )/\mathcal{N}_{\mu}(\bm{k}_{l}).
\end{nalign}
Here as before $\mathcal{N}_{\mu}(\bm{k}_{l}) = |\det \braket{\psi(\bm{k}_{l})|\psi(\bm{k}_{l}+\hat{\mu})} |$, and $\psi(\bm{k}_{l})$ denotes the multiplet $(\ket{n_{1}}, \ldots ,\ket{n_{m}})$, where $m$ is the largest degeneracy in the occupied bands. 

If we have a mirror symmetry $M_{i}$ for a Bloch Hamiltonian $h(\bm{k})$ that satisfies $M_{i}h(\bm{k})M_{i}^{-1} = h(-k_{i},\tilde{\bm{k}})$, where $\tilde{\bm{k}}$ denotes the momenta unchanged by $M_{i}$, then, in the plane $k_{i}= 0$, $h$ commutes with $M_{i}$. $h$ is then block-diagonal in the two $M_{i}$ sectors (labeled by $\pm i$); i.e., $h  = h^{+}\oplus h^{-}$. With this definition, a Chern number $C^{+/-}$ may be computed for each sector. Furthermore, a mirror Chern number defined for the plane is given by \cite{Teo2008a} 
\begin{nalign}
n_{M_{i}} = (C^{+}-C^{-})/2.
\end{nalign} 
Using $n_{M_{i}}$ gives the number of gapless modes along the mirror plane boundary, as seen in the main text.
\section{ Minimal Dirac Hamiltonian Method}\label{app:MDH}
Here we give a brief overview of the minimal Dirac Hamiltonian method used to determine the topological classification under the mirror symmetry \cite{Chiu2013}. Note that we keep the mirror symmetry representation such that the eigenvalues are $  \pm i$, in contrast to \cite{Chiu2013}. 

We write a Hamiltonian in $d$ spatial dimensions as 
\begin{nalign}
H = m \gamma_{0} + \sum_{i=1}^{d} k_{i}\gamma_{i}.
\end{nalign} 
Here $m$ is constant and we have the commutation and anticommutation relations 
\begin{nalign}
\{\gamma_{i},\gamma_{j}\} = 2\delta_{ij}\mathbb{I}, \ i = 0,1,\ldots, d, \\
\begin{array} { c } { \left[ \gamma _ { 0 } , T \right] = 0 , \quad \left\{ \gamma _ { i \neq 0 } , T \right\} = 0, } \\ { \left\{ \gamma _ { 0 } , C \right\} = 0 , \quad \left[ \gamma _ { i \neq 0 } , C \right] = 0 ,}  \end{array} \end{nalign}
where $T, C$ denote time-reversal symmetry (TRS) and particle-hole symmetry (PHS), respectively. 
If it is possible to write an extra mass term to be added into the Hamiltonian such that: the new mass term anticommutes with $\gamma_{0}$, respects the given symmetries, and opens a gap in the spectrum of the system such that varying $m$ in the mass term $m\gamma_{0}$ keeps the system in the same topological phase, then we denote this term as a symmetry-preserving extra mass term (SPEMT). If no such SPEMT term exists, then the system possesses either $\mathbb{Z}_{2}$ or $\mathbb{Z}$ topology. To differentiate between $\mathbb{Z}_{2}, \mathbb{Z}$, we consider an enlarged Hamiltonian 
\begin{nalign}
H' = \sum_{i} k_{n_{i}}\gamma_{n_{i}}\otimes \sigma_{z} + \sum_{\text{remain}}k_{n_{j}}\gamma_{n_{j}}\otimes \mathbb{I},
\end{nalign} 
where $n_{i} \in (0,1,\ldots, d)$. The second summation is over the gamma matrices not included in the first summation. If a SPEMT can be added to this larger minimal Dirac Hamiltonian, then the topological classification is $\mathbb{Z}_{2}$.

To find the topological classification under mirror symmetry, we construct an operator $\gamma_{1}R$ satisfying 
\begin{nalign}
\{\gamma_{1}R,H\} = 0
\end{nalign} 
The above operator may be added either as an extra mass term or an extra kinetic term, depending on the indices $\eta_{T},\eta_{C}$ stated in the main text. The addition of this operator as an extra kinetic term increases the effective dimension of the Hamiltonian, while addition of $\gamma_{1}R$ as an extra mass term decreases the effective dimension of the Hamiltonian. The topological invariants for these higher (lower) dimensional Hamiltonians without reflection symmetry are in correspondence with the topological invariants of the Hamiltonian with the reflection symmetry.
%\bibliography{WSMTSC}

\end{document}